# Grouping environmental factors influencing individual decision-making behavior in software projects: a cluster analysis

Jingdong Jia[1*], Hanlin Mo[2], Luiz Fernando Capretz[3] and Zupeng Chen[1]

[1] School of Software, Beihang University, Beijing, 100191, China

[2] Institute of Computing Technology, Chinese Academy of Sciences, Beijing, 100190, China

[3] Department of Electrical & Computer Engineering, Western University, London, Ontario, N6A5B9, Canada

**Abstract**:

An individual's decision-making behavior is heavily influenced by and adapted to external environmental factors. Given that software development is a human-centered activity, individual decision-making behavior may affect the software project quality. Although environmental factors affecting decision-making behavior in software projects have been identified in prior literature, there is not yet an objective and a full taxonomy of these factors. Thus, it is not trivial to manage these complex and diverse factors. To address this deficiency, we first design a semantic similarity algorithm between words by utilizing the synonymy and hypernymy relationships in WordNet. Further, we propose a method to measure semantic similarity between phrases and apply it into k-means clustering algorithm to group these factors. Subsequently, we obtain a taxonomy of the environmental factors affecting individual decision-making behavior in software projects, which includes eleven broad categories, each containing two to five sub-categories. The taxonomy presented herein is obtained by an objective method, and quite comprehensive, with appropriate references provided. The taxonomy holds significant value for researchers and practitioners; it can help them to better understand the major aspects of environmental factors, also to predict and guide the behavior of individuals during decision making towards a successful completion of software projects.

## 1. Introduction

During software development and evolution, many decisions have to be made concerning people, processes, techniques and tools.[1] For example, facing conflicts, some developers choose a positive cooperative strategy, but others choose a negative avoidance strategy. What makes software engineers choose different decision-making behaviors? Different decision-making behaviors bring different outcomes, so the project's success depends on how individuals in a team deal with problems and make decisions. Although many tools and techniques (for example, checklists and decision models) have been proposed to assist decision-making in software projects, no one can explain how people make decisions in their contexts.[1] Individual decision-making behavior is still seldom focused on in software projects, although it is important for software development and evolution. Recently, Lenberg et al.[2] proposed a concept of behavioral software engineering by taking cues from behavioral economics, which underpins the research that focuses on behavioral aspects of software engineers.

According to social cognitive theory, overt behavior is influenced by intrinsic personal factors and extrinsic environmental factors.[3,4] Therefore, the factors influencing individual decision-making behavior include two aspects: personal and environmental factors. Personal factors, especially personality, have been explored a lot in the software engineering (SE) domain. Environmental factors are also known as situational or contextual factors. There exists some literature discussing environmental factors with regard to some topics; for example, environmental factors influencing IT personnel's intentions to leave,[5] software development process[6] and employee self-development.[7] However, environmental factors are different for different issues. Little focus, though, has been placed on environmental factors influencing individual decision-making behavior, which actually have an important effect on software development, maintenance and evolution.

In fact, environmental factors are important for decision making because decisions are not made in isolation,[8] but rather in several different contexts. Software development process depends on the situational characteristics. Such characteristics include the nature of the application(s) under development, team size, requirements volatility and personnel experience.[6] The rationale of decisions may be influenced by external forces and constraints.[9] Therefore, when making a decision, individuals in a software team should always consider environmental factors, which can affect the decisions themselves as well as potential consequences of decisions even in environments that are relatively stable. For example, design decisions in an agile environment are different from that in an organization with strictly defined hierarchies following a plan-driven process. In addition, we all know that technology can streamline the process of project in many cases. The decision involving which technology will be used in a software project depends on many environmental factors; the necessary technology may be expensive or not compatible with existing technology or equipment. Even when technology is available, in a limited software development time, the fact that individuals are not familiar with the technology also influences their adoption of that technology. Therefore, the impact of environment factors on individual decision-making cannot be ignored. Otherwise, poor, even bad decisions are made causing project delays and failures.[9] For example, if individuals choose a new but unfamiliar technology in a limited time, then training may become an issue and can create delays and add expense for projects. Unanticipated changes in the environment can cause even the most well-managed and smoothly proceeding project to lose momentum. Therefore, it is imperative to pay attention to environmental factors that affect the decisions. Exploring how different context-related factors influence the decision-making process can help to identify best practices for making decisions.[8]

This study is the based on a prior study,[11] in which eight categories and 237 environmental factors that affect individual decision-making behavior were identified by systematic literature review (SLR). That work, however, has some room for improvement. Firstly, the eight categories were extracted only from eight papers, but the 237 environmental factors came from 40 pieces of literature. The prior work[11] has mentioned that their categories may not cover all the factors. So, whether or not the eight categories are appropriate to describe these environmental factors is up for debate. Additionally, classifying each factor into a category in their work relied on subjective assessments of semantic relatedness between the factor and a category. The subjective assessment method may affect the classification result. Consequently, the number of factors in some categories is too big (for example, 58 or 52), and other too small (for example, 5). And, for these categories including many environmental factors, their work did not provide further subdivision, so it is not easy to understand and use these big categories. In existing literature,[6,8] many factors were usually organized in a hierarchy with two levels. Therefore, in this study, we aim to expand the classification and use an objective method to provide a more detailed taxonomy of these environmental factors influencing individual decision-making behavior in software projects.

We address the following three research questions:

RQ1 - How many categories are appropriate to classify these 237 environmental factors?

RQ2 - Which word or phrase can accurately describe the meaning of each category?

RQ3 - Which category does each environmental factor belong to?

To answer these question, we adopt a cluster analysis to form a taxonomy of 237 environmental factors provided in the study [11] that are our dataset. We believe clustering can reveal the very complex relationship between objects and features,[12] and was widely use to solve the correlation problem, for example, in the study by Choi et al.[13] For our research problem, semantic clustering is suitable because the objective of this clustering technique is grouping elements based on the similarity in their content.[14] Common clustering algorithms depend on choosing a similarity measure between data points and a correct clustering result can be dependent on an appropriate choice of a similarity measure. The choice of a correct measure must be defined relative to a particular application.[15] In this sense, the assessment of semantic similarity between environmental factors is a key task in our research. Semantic similarity states how taxonomically close two terms are, because they share some aspects of their meanings.[16] Each environmental factor is either a word or a phrase. In fact, a word is a special phrase whose length is 1. Therefore, we propose a method to measure semantic similarity between words based on the famous electronic lexical database: WordNet. Further, we design a semantic similarity algorithm for phrases. Based on the method of semantic similarity, we choose a clustering algorithm to cluster these environmental factors.

The final goal is to provide an objective and detailed taxonomy of environmental factors influencing individual decision-making behavior, so that researchers and practitioners can better understand and predict the individual decision-making behavior, and design more effective solutions to improve personnel management in SE.

This paper is organized as follows. Section 2 describes the related work. Section 3 presents our research method. Section 4 provides the analysis of clustering results. We subsequently discuss the results in terms of their theoretical and practical contributions in Section 5. Limitations and future study are discussed in Section 6. Our conclusions are drawn in Section 7.

## 2. Related work

In this section, we briefly overview the results which are most relevant for the present work.

### 2.1 Word clustering

Word clustering refers to the process of partitioning a collection of words into several subsets, called clusters, so that clusters exhibit high intra-cluster and low inter-cluster similarity;[17] the words within the same cluster are similar to each other and, simultaneously, dissimilar to words in the other clusters.[18,19] Word clustering has been one of the most challenging tasks in natural language processing.[20] It has many direct and relevant applications. For example, it is particularly useful in automatic categorization of texts or documents,[12,18,21-23] news article clustering,[24] and web sentence retrieval.[25] In addition, word clustering is also widely used to group words in a specific professional domain; for example, biology[26] and medicine.[27] Our research focuses on word clustering related to the field of SE.

There have been a number of methods proposed in the literature to address the word clustering problem.[15] Word clustering relies on three concepts: a representation of word semantic features, a similarity measure and a clustering algorithm that builds the clusters using the word feature data and the similarity measure.[17] Getting and expressing the semantic features of a word is fundamental, which provides prerequisite for measuring the semantic similarity. In general, the semantic, syntactic and statistical properties of words can be utilized to capture features of words in various word clustering algorithms.[20,25] The syntactic information of words focuses on how words are organized in a sentence. Statistical information generally refers to the probability of the occurrence of a word in a given context. Contextual information in a certain corpus is often used as the basic word feature type.[28,29] It is a reasonable assumption that words occurring in the same context tend to have similar meaning. Therefore, words are clustered according to their frequencies in the context; words having similar co-occurrence patterns are classified in the same class.[15] Additionally, the semantic information of words, that is the meaning of the words themselves, is also used to improve the quality of clustering in literature.[21,30]

In our research, we extract word features based on their semantic information to calculate their semantic similarity for two reasons. First, our dataset is a collection of words or phrases, which do not provide enough contextual and syntactic information. In addition, our goal is to form a taxonomy in which words in the same cluster represent one theme of environmental factors, so word meanings are better than word forms that refer to the occurrences of words in a corpus. This is also in line with the idea in a study by Li, Chung and Holt .[21] The semantic information of words is extracted from WordNet, which is discussed in the next Section. The clustering algorithm we chose is the k-means algorithm because it is a well-known and popular partitioning method for clustering,[31] where clusters are identified by minimizing the clustering error.[32]

**2.2 WordNet**

WordNet is one of the most widely used and largest electronic lexical databases of English.[30,33] It was originally developed by scientists at Princeton University in the 1990s and continues to be developed and maintained to provide an online dictionary constructed not merely in alphabetical order but in a more conceptual way showing semantic relationships among concepts.[34] In WordNet, word forms include nouns, verbs, adjectives and adverbs, of which nouns are first developed and the most mature.

WordNet provides many types of relationships among concepts.[35] First, synonymy is WordNet's basic relation, because WordNet groups words into sets of synonyms called synsets, each expressing a distinct concept. The synsets are organized into senses, giving thus the synonyms of each word.[33] Synonymy is a symmetric relation between word forms. Additionally, synsets are interlinked by means of conceptual-semantic and lexical relations. Hypernym relationships, and its inverse, hyponymy, are important transitive relations between synsets. Because there is usually only one direct hypernym, hypo/hypernym relationships provide a hierarchical tree-like structure for each term.[30,35]

Due to the full semantic features of words, WordNet has been widely used in many studies for different purposes, for example, Lee et al.[34] proposed the automatic generation of concept hierarchies using WordNet. However, the dominant research domain is to calculate the similarity between words or concepts,[35] or improve the accuracy of clustering techniques related to words.[30] Based on WordNet, different methods to determine similarity between terms have been proposed to solve different problems, for example, Zhu et al.[36] presented a method for measuring the semantic similarity between concepts in knowledge graphs.

### 2.3 Individual decision-making and software projects

Individual decision-making has been investigated over decades in several disciplines, for instance, economics and social psychology.[37] However, there has been very limited research on individual decision-making in SE field. Many types of activities in software development involve decisions that have significant consequences on the process and the final product.[38] As a result, there is a strong call and need to examine individual decision-making in software projects.

The decision-making of a special individual (software project manager) was conducted in order to increase the effectiveness of software project management.[10] In addition, individuals make decisions in different stages of software projects. So, individual decision-making in requirements engineering was focused from different research perspectives.[39,40] How developers make design decision was examined,[41] and developers' design decision in agile was compared with non-agile project.[43] Because developers in open-source software projects are typically not co-located and not everyone works in the same company, how they make unified architectural decisions has been discussed.[9] Developers' decisions made in an iteration cycle were examined and summarized, and the obstacles to decision making in agile process were identified.[43]

Regarding individual decision-making for certain things in software projects, Børte et al.[44] investigated how software professionals reach a decision on software effort estimate. Developers' decision about whether report bad news on software projects or not has been examined.[45] Hahn et al.[46] investigated how individuals make decisions about which teams to join in the context of open source software development. Developers' decision-making about the evolution of Python language were discussed.[47] And Hirao et al.[48] investigated the method of reviewer's decisions on the code review.

### 2.4 Environmental factors and software projects

Our ability to improve decision making in software development hinges on our understanding how decisions are made.[42] Therefore, in addition to knowing the decision contents, methods and processes, we also need to know which factors affect individual decision-making. Obviously, personal factors may affect individual decision-making, for example, personal preferences[49], individual experiences[42], and personal personality and knowledge.[10] However, in software projects, individual decision-making is a complex process, and also depends on environmental factors. Although some studies have involved this aspect, which environmental factors influence individual decision-making has not yet been fully addressed.

Exploring how different context-related factors influence the decision-making process can help to identify best practices for making decisions.[8] In the work of Cunha et al.,[10] eight situational factors related to the decision of software project manager were identified: autonomy of software project manager, constant feedback, client involvement, support departments involvement, iterative planning, knowledge sharing initiatives, team members' commitment and technical capacity. Harrison et al.[9] said that the rationale of decisions may be influenced by external forces and constraints, such as technology. A method to analyze environmental factors that cause stakeholders' decision about requirements changes was provided,[50] but there were no detailed environmental factors. In the work of Groher and Weinreich,[8] seven categories of context-related factors were given, among which three categories (company size, cultural factors, decision scope) have not sub-factors, and other four categories (project factors, business factors, organizational factors and technical factors) have total 12 sub-factors. More environmental factors in software projects were listed by a SLR method,[11] but the classification about these factors is not sufficient.

## 3. Research method

We address the issue of determining a taxonomy of environmental factors influencing individual decision-making behavior in software projects through cluster analysis. The issue includes two aspects: how many categories are appropriate and what elements are included in each category? Additionally, we want to find a suitable factor to represent each category. In order to resolve our problems, we propose a research method including four steps, as shown in Figure 1.

<< Figure 1 >>

**Figure 1** Overview of the research method

### 3.1 Normalizing the data

In the literature,[11] 237 environmental factors were provided in a table. We observed that these factors were single words or short phrases. In this work, we regard each environmental factor as a phrase. Additionally, taking parts of speech into perspective, those single words are nouns, for example, autonomy and communication, which are the first and third major environmental factors in the study.[11] Also, these short phrases are mostly noun phrases, in which the head is a noun optionally accompanied by a set of modifiers used to identify it in detail,[22] for example, good management and appropriate working conditions. Although WordNet includes nouns, verbs, adjectives and adverbs, the noun section is the most developed. Taking the characteristics of our data and WordNet into account, we focus only on nouns. Thus, we first normalized the original word in each phrase into the corresponding singular noun.

Obviously, some words, for example a conjunction, an article or a preposition in a phrase do not have a corresponding noun form. In this case, we removed them from the phrase. Although these words reflect the syntactic construction of phrases, they actually don't have a semantic meaning for clustering analysis. In addition, the proportion of them in our dataset is not high; there were total 532 words in all environmental factors, and 420 words have the corresponding nouns. Nouns account for 79%. The high proportion of nouns shows that it is reasonable to only consider words with the corresponding nouns for cluster analysis.

After normalizing the data, the new dataset, which is denoted as $S_{ef}$, still includes 237 elements that were noun phrases of environmental factors. The subsequent analysis is based on $S_{ef}$.

**3.2 Designing semantic similarity algorithm for words**

As mentioned before, classification of the environmental factors is based on their semantic similarity; factors with similar meaning belong to the same category. Although all factors were expressed in phrases, words are the basis of a phrase, thus it is necessary to first calculate the degree of semantic similarity between two words.

From an information theory point of view, the similarity between two objects is regarded as how much they share in common.[17] In the real world, people may use different word forms to express the same word meaning, and those word forms are called synonyms. A word meaning can be represented by a synonym set, a set of word forms that are synonyms.[21] Synonymy is the primary semantic relationship between words we used for clustering. Most existing WordNet-based clustering methods utilize synonymy to identify semantically similar concepts, for example, a study of Zheng and Kang.[22] Additionally, hypo/hypernymy, an important semantic relationship between words, represents the connection between a specific and a general word meaning.[21] Using hypernyms can help magnify hidden similarities to identify related topics, which potentially leads to clustering quality,[19] so the hypernyms of WordNet have been used to explore the semantic relations between terms.

Therefore, we design the semantic similarity algorithm for words by exploring the synonymy and hypernymy relationships between words based on WordNet.

**3.2.1 Measuring semantic similarity between words based on synonymy relationship**

WordNet contains the senses of a word, which are really the meanings of the word. Generally, in WordNet, a word may have multiple senses, which are ordered from the most to least frequently used. Additionally, for each sense of a word, a set of synonyms that includes the word itself is given. Suppose $w_a$ and $w_b$ are any two words, and they separately have multiple senses. We compare each sense of $w_a$ with every sense of $w_b$ to measure the semantic similarity of them by utilizing the information of synonyms in WordNet. We define a variable $S_s(w_a, w_b)$ to represent the semantic similarity between the two words based on a synonymy relationship. The detailed algorithm of $S_s(w_a, w_b)$ is given below.

Because a word may have multiple senses, we use a superscript to denote its different senses; $w_a^i$ means the $i$th sense of $w_a$, and $w_b^j$ means the $j$th sense of $w_b$. We first measure the semantic similarity of $w_a^i$ and $w_b^j$. We use $SY_{w_a^i}$ to represent the set of synonyms of $w_a$ under its $i$th sense. Similarly, $SY_{w_b^j}$ represents the set of synonyms of $w_b$ under its $j$th sense. If the intersection of $SY_{w_a^i}$ and $SY_{w_b^j}$ is not empty, it is reasonable to think that the two words, $w_a$ and $w_b$, have the semantic similarity under this pair of senses. Otherwise, their semantic similarity does not exist. We define a variable $g(w_a^i, w_b^j)$ to represent whether the semantic similarity of $w_a^i$ and $w_b^j$ exists. The value of $g(w_a^i, w_b^j)$ is assigned with the following equation.

$$g(w_a^i, w_b^j) = \begin{cases} 1, & if \ SY_{w_a^i} \cap SY_{w_b^j} \neq \emptyset \\ 0, & elsewise \end{cases} \quad (1)$$

Because each sense of a word has different frequency of usage, we introduce $r_i$ ($0 \leq r_i \leq 1$) as the weight of the $i$th sense of any word to reflect the fact that the importance of every sense of a word is different. Because the multiple senses of a word are ordered from the most to the least frequently used in WordNet, the value of $r_i$ is descending from $r_1$. Thus, $r_i * r_j * g(w_a^i, w_b^j)$ measures the degree of semantic similarity of $w_a^i$ and $w_b^j$.

Although a word may have multiple senses, we select only the first three because the percentage of their usage frequency is highest. In the literature,[21] only the first two senses were selected. In addition, there are total over 400 words in our dataset $S_{ef}$. Taking the running time and the accuracy of the algorithm into account, we consider it reasonable to only use the first three senses to calculate semantic similarity of words in this investigation. If the number of senses of a word is less than three, we use all its senses in our algorithm. Based on multiple experiments, the weights of three senses are set as: $r_1$=0.6, $r_2$=0.3, and $r_3$=0.1.

To calculate $S_s(w_a, w_b)$, we compare the semantic similarity of each selected sense pair of $w_a$ and $w_b$, then sum them. Therefore, $S_s(w_a, w_b)$ can be represented as follows:

$$S_s(w_a, w_b) = \sum_{i=1}^{3} \sum_{j=1}^{3} r_i * r_j * g(w_a^i, w_b^j) \tag{2}$$

### 3.2.2 Measuring semantic similarity between words based on hypernym relationship

For each sense of a word, WordNet provides additional semantic hierarchies; hypernym relations,[34] which are an important indicator of depth of knowledge.[35] We define a variable $S_h(w_a, w_b)$ to represent the semantic similarity between two words, $w_a$ and $w_b$, based on their hypernym relationship. In WordNet, for each sense of a word, there are multiple levels of hypernymy. In order to improve the time efficiency of our algorithm designed to calculate $S_h(w_a, w_b)$, we only navigate upwards for the direct five levels of hypernymy.

Here, we use two superscripts to denote its different senses and hypernymy; $w_a^{il}$ means the $l$th ($1 \leq l \leq 5$) level of hypernym of the word $w_a$ under the $i$th sense, and $w_b^{jf}$ means the $f$th ($1 \leq f \leq 5$) level of hypernym of the word $w_b$ under the $j$th sense. We first measure the semantic similarity of $w_a^{il}$ and $w_b^{jf}$. For any word, its hypernym in a certain level may also be more than one in WordNet. Let $H_{w_a^{il}}$ denote the set of hypernyms of $w_a^{il}$. Similarly, $H_{w_b^{jf}}$ denote the set of hypernyms of $w_b^{jf}$. If the intersection of $H_{w_a^{il}}$ and $H_{w_b^{jf}}$ is not empty, it means that $w_a^i$ and $w_b^j$ have a semantic relationship in this level pair of hypernyms. We define a variable $h(w_a^{i_l}, w_b^{j_f})$ to represent whether the relatedness of $w_a^{il}$ and $w_b^{jf}$ exists, and it is expressed as follows:

$$h(w_a^{i_l}, w_b^{j_f}) = \begin{cases} 1, if \ H_{w_a^{i_l}} \cap H_{w_b^{j_f}} \neq \emptyset \\ 0, \qquad\qquad elsewise \end{cases}$$

(3)

Because hypernyms are given in a hierarchical structure, the higher level means a longer path. Generally, the shorter the path from one node to another, the more similar they are. Therefore, we introduce a variable *dep* to describe the distance of two level, and let *dep=|l-f|*. Obviously, the value of *dep* is 0, 1, 2, 3, or 4 because the values of *l* and *f* are integers from 1 to 5. In order to reflect the influence of distance on the similarity, we set $u_{dep}$ a distance constant coefficient, which is a decimal in a descending order from $u_0$ to $u_4$. Thus, the semantic similarity of $w_a^{i_l}$ and $w_b^{j_f}$ is $u_{|l-f|} * h(w_a^{i_l}, w_b^{j_f})$. In experiments, we set $u_0 = 0.5$, $u_1 = 0.25$, $u_2 = 0.125$, $u_3 = 0.0625$, and $u_4 = 0.03125$.

Because the hypernym relationship is less semantically similar than the synonym relationship, and there are 25 pairs $h(w_a^{i_l}, w_b^{j_f})$ when *l* and *f* change from 1 to 5, for simplicity, we choose the maximum of the semantic similarity of $w_a^{i_l}$ and $w_b^{j_f}$ as the semantic similarity of $w_a^i$ and $w_b^j$, which is expressed as $S_h(w_a^i, w_b^j)$ in the following equation:

$$S_h(w_a^i, w_b^j) = \max\{u_{|l-f|} * h(w_a^{i_l}, w_b^{j_f})\}, 1 \leq l, f \leq 5 \qquad (4)$$

For $w_a$ and $w_b$, considering their different senses, the semantic similarity for the hypernym relationship of them, expressed as $S_h(w_a, w_b)$, can be calculated as follows:

$$S_h(w_a, w_b) = \sum_{i=1}^{3}\sum_{j=1}^{3} r_i * r_j * S_h(w_a^i, w_b^j) = \sum_{i=1}^{3}\sum_{j=1}^{3} r_i * r_j * \max\{u_{|l-f|} * h(w_a^{i_l}, w_b^{j_f})\}, 1 \leq l, f \leq 5 \quad (5)$$

In equation (5), we also only consider the first three senses of two words to improve the running efficiency of the algorithm.

### 3.2.3 Measuring semantic similarity between words

We have gotten the semantic similarity of two words from two aspects: $S_s(w_a, w_b)$ and $S_h(w_a, w_b)$, and we think the two parts have the same contribution to the semantic similarity of two words. Thus, we assign them equal weight: 0.5. Then, to simplify the algorithm, the simplest addition principle is adopted to measure the semantic similarity of two words, $w_a$ and $w_b$, which is expressed as $S(w_a, w_b)$. Therefore, the final equation of the degree of semantic similarity of two words, $S(w_a, w_b)$, is the following:

$$S(w_a, w_b) = 0.5 * S_s(w_a, w_b) + 0.5 * S_h(w_a, w_b) \qquad (6)$$

### 3.3 Designing semantic similarity algorithm for phrases

The elements in our dataset $S_{ef}$ are all phrases including at least one word. In this section, we propose a semantic similarity algorithm for phrases in detail, which is based on the algorithm provided in Section 3.2.

For any two phrases *A* and *B*, we use $PhraseS(A, B)$ to denote the semantic similarity of them. Without losing generality, we assume *A* and *B* separately contain *m* and *n* words, and, obviously, *m* and *n* are both bigger than 0. We use *A*[*x*] and *B*[*y*] to represent the *x*th word in *A* and the *y*th word in *B*. According to equation (6), we can calculate the semantic similarity of *A*[*x*] and *B*[*y*], $S(A[x], B[y])$. If the semantic similarity of two phrases is high, it is reasonable to assume that the semantic similarity of some words in the two phrases are high. Therefore, we directly calculate the semantic similarity between each word in *A* and every word in *B*. Because *x* ranges from 1 to *m*, and *y* ranges from 1 to *n*, we can get total *m*\*n* semantic similarity of word pairs between *A* and *B*. Then, we sum them. Additionally, in order to keep the value of $PhraseS(A, B)$ in the range of [0, 1], we define the final equation of $PhraseS(A, B)$ as follows:

$$PhraseS(A, B) = \left(\sum_{x=1}^{m} \sum_{y=1}^{n} S(A[x], B[y])\right) / m * n \qquad (7)$$

In our dataset, we have 237 phrases of environmental factors. Using equation (7), we can calculate the semantic similarity of any two environmental factors. Thus, we get a 237×237 semantic similarity matrix M, in which $M_{ij}$ represents the semantic similarity of the *i*th environmental factor and *j*th environmental factor. The matrix M is the base of the clustering analysis in the next section.

### 3.4 Clustering analysis based on k-means

After designing the algorithm to measure the semantic similarity between phrases, we use the k-means algorithm to cluster these phrases in order to find the suitable classification of environmental factors, because this remains one of the most popular methods. In fact, it has been identified as one of the top 10 algorithms in data mining.[51]

A given dataset is grouped into a predetermined number *k* of disjoint sets, called clusters, through k-means. The user must specify the desired number of clusters, *k*, before the clustering process. However, with regards to our research, it is difficult to specify a reasonable number of clusters because we have so little information. In fact, how many categories are appropriate for these environmental factors is one of our research questions. Instead of telling the number of clusters to the clustering algorithm, it makes more sense to let the clustering algorithm itself find out *k*. Therefore, we may run the k-means algorithm several times with different number of clusters in a reasonable range in order to choose an appropriate one, although this is time-consuming.

To choose an optimal number of clusters, we design several indices to compare the clustering results under different number of clusters. The aim of our clustering is to make the intra-cluster semantic similarity high and the inter-cluster semantic similarity low as soon as possible. Thus, we define an index: the semantic similarity of a cluster. In the following, we describe how to calculate the index.

Suppose our dataset was grouped into *k* clusters {$c_1$, $c_2$,…, $c_k$} through k-means algorithm. The cluster $c_i$ includes *m* elements with a centroid phrase $d_i$. We define a variable $S_{ci}$ to describe the semantic similarity of this cluster $c_i$. It is equal to the sum of the semantic similarity between any phrase in this cluster and the centroid phrase, $d_i$, divided by the number of phrases contained in the cluster. According to equation (7), therefore, the equation of $S_{ci}$ is as follows:

$$S_{ci} = \sum_{j=1}^{m} PhraseS(A_j, d_i) / m \qquad A_j \in c_i \qquad (8)$$

Based on equation (8), obviously, under a given k, we can calculate the maximum, minimum and the average of the semantic similarity of *k* clusters, which are just our comparison indices, and expressed in the following three equations.

$$S_{kmax} = \max\{S_{ci}, i = 1,2,\cdots,k\} \qquad (9)$$

$$S_{kmin} = \min\{S_{ci}, i = 1,2,\cdots,k\} \qquad (10)$$

$$S_{kavg} = \sum_{i}^{k} S_{ci}/k, i = 1,2,\cdots,k \qquad (11)$$

Then, we describe how to apply the k-means algorithm to cluster environmental factors. This algorithm has two stages: initialization, in which we set the starting set of centroids, and iteration.[51] A classic k-means algorithm iteratively performs clustering until the desired number of clusters, *k*, is obtained.[28] Since the best number of clusters is unknown prior to clustering for our research, as mentioned before, we set a range of the number of clusters: an integer from *m* to *n*. For a certain *k* in the range, we describe the detailed steps of k-means in the following.

Step 1: setting the initial *k* centroids.

This is the initialization stage. Some studies have stated that the initialization would affect the quality of clustering; a poor initialization could lead to a poor local optimal.[51] Therefore, we try to make our initialization more reasonable. We describe how we gained our initial centroids in the following.

According to the number of our dataset $|S_{ef}|$ and the number of clusters *k*, in order to balance the number of factors in each original cluster, first, we calculated $|S_{ef}|/k$, and round it into an integer, which is denoted as $int(|S_{ef}|/k)$. Then, in order to obtain initial centroids, we construct initial *k* clusters denoted by $\{c_1, c_2,\ldots, c_k\}$; from the first environmental factor in dataset, every $int(|S_{ef}|/k)$ elements are grouped into a cluster. For any $c_i$, suppose $p_j$ is any phrase in $c_i$. We use $|p_j^{0.2}|$ to denote the number of phrases in $c_i$, whose semantic similarities with respect to $p_j$ are bigger than the threshold value 0.2. Based on the semantic similarity matrix M obtained in Section 3.3, we can obtain the values of $|p_j^{0.2}|$ for any $p_j$.

Then, we find the maximum of them. If $t = arg \max|p_j^{0.2}|, p_j \in c_i$, then $p_t$ that corresponds to the maximum $|p_j^{0.2}|$ is chosen as the centroid of this cluster, that is $d_i = p_t$. For each cluster, we do the same operation. Thus, we can get *k* initial centroids $d_1, d_2,\ldots d_k$.

The next two steps belong to the iterative stage.

Step 2: Determining the elements of every cluster according to centroids

For each iteration, we first determine which cluster each phrase belongs to. The general principle is that the semantic similarity between a phrase and a centroid should be as big as possible. Therefore, for any phrase *E* in $S_{ef}$, we calculate the semantic similarity between *E* and *k* centroids based on the equation (7). Then, we find out the centroid $d_{t'}$ that has the biggest semantic similarity with respect to *E*; $t' = arg \max(PhraseS(E, d_i)), i = 1,2,\cdots k$. Thus, the phrase *E* is in the cluster $d_{t'}$. According to this method, each phrase is grouped into a cluster, and we can obtain all the elements of each cluster.

Step 3: Updating centroids for each cluster during each iteration.

Once we get *k* clusters and their elements, the centroid of each cluster is updated. For each element, $A_i$, of a cluster *c*, we calculate the sum of the semantic similarity between $A_i$ and other phrases in the same cluster. In other words, according to equation (7), for $A_i$, we calculate $\sum_{j=1}^{|c|} PhraseS(A_j, A_i), A_j \in c$. Then we find the maximum sum, and the corresponding $A_i$ is set as the new centroid of this cluster.

Steps 2 and 3 are repeated until an iteration stopping criterion is met. Generally, one stopping criterion is a prespecified number of iterations. After multiple experiments, taking running time of algorithm into account, in our work we set the number of iterations as 15. In addition, the algorithm stops if the centroids of each iteration do not change any more.

## 4. Results and analysis

In our experiments, we need to set the range of the optimal number of categories. Because there are many environmental factors that are clustered, obviously, if the number of categories is set very small, the semantic similarity of each cluster is lower, and the clustering results have little meaning for our research. If the number of categories is set very big, it is not beneficial to manage so many factors. Thus, the number of categories is set as an integer from 5 to 12. Using the method presented in Section 3.4, we obtain the answers to the research questions, which are provided here.

### 4.1 The number of categories (RQ1)

In order to find the optimal number of categories, we run the clustering algorithm provided in Section 3.4 eight times when the number of categories, $k$, changes from 5 to 12. Under a given $k$, the maximum, minimum and the average of the semantic similarity of $k$ clusters are found and listed in Table 1.

<<Table 1>>

According to our research question, obviously, cases where the semantic similarity of a cluster is greater are better, because it means the environmental factors in a cluster are highly semantically similar. We use three indices, $S_{kmax}$, $S_{kmin}$, and $S_{kavg}$, to compare the results of different $k$ so as to choose the most suitable number. Ideally, if the results of the three indices under a certain number $k$ are all bigger than the corresponding values under other numbers, it is easy to know that the optimal number is just the $k$. However, the ideal condition did not happen in our results. From Table 1, we can see that $S_{kmax}$ is biggest when k is equal to 10, $S_{kmin}$ is biggest when k is equal to 11, and $S_{kavg}$ is biggest when k is equal to 12. The three indices provide different results. In this case, from the viewpoint of statistics, it is reasonable to choose the optimal number of categories according to the average value, $S_{kavg}$. Thus, the case that k=12 is better than that under k=11. However, we can see the difference of average values is very little under two cases. In addition, if the results under k=12 is better than that under k=11, we have a reason to doubt whether it is better to make the number of categories much bigger. Therefore, we changed $k$ from 13 to 20, and conducted clustering to test the idea. The average semantic similarity of clusters ($S_{kavg}$) when $k$ changes from 5 to 20 is shown in Figure 2.

From Figure 2, we can see that the average semantic similarity increases significantly when the number of categories changes from 5 to 10. Then, it tends to be stable, especially when the number of categories is bigger than 12, which indicates that the upper limit of $k$ that we have set is reasonable.

<< Figure 2 >>

**Figure 2** The average semantic similarity of clusters under different numbers of categories

In addition, from Figure 2, we can see that the difference of $S_{kavg}$ under three cases (the number of categories is 10, 11, or 12) is not significant. Combined with Table 1, we find that the minimum semantic similarity ($S_{kmin}$) is best when $k=11$, which means that every cluster has good similarity. In addition, in this case of $k=11$, its maximum and average semantic similarity are separately only a little smaller than when $k=10$ and $k=12$. Therefore, we think the optimal number of categories of these environmental factors is 11. In other words, it is appropriate to group these environmental factors into eleven categories.

### 4.2 The name and number of each category (RQ2)

Once the optimal number of categories is determined, the answer to RQ2 is also obtained according to our clustering algorithm. The centroids are regarded as the names of categories to describe the meaning of each category. Table 2 shows the name of each category, which answer the research question RQ2, and the total number of elements in each category.

<<Table 2>>

The environmental factors have been classified into eight categories that are subjectively obtained by SLR in a study.[11] Compared with their results, we find that four categories, which are italicized in Table 2, are almost the same. Therefore, task, team, organization and competence are four important categories that should be paid attention to. In addition, from the viewpoint of the number of elements in each category, the number of elements in the four categories in our research are all less than the results in the previous study.[11] The four categories are the major categories in the study,[11] but they are not in our results. We think that the reason for the difference is that our eleven categories are more subdivided.

As mentioned in Section 1, we have stated that the number of elements in some categories of the study[11] is too big to manage these categories. Further, we compare the maximum and minimum number of elements in our result with their study, as shown in Figure 3. Obviously, the difference between the maximum and the minimum number of elements in literature[11] is greater than that of our study, which indicates that our classification is relatively more balanced than their subjective classification.

<< Figure 3 >>

**Figure 3** The comparison of the maximum and minimum number of elements in two studies

### 4.3 Taxonomy of environmental factors (RQ3)

When we chose 11 categories, which category each environmental factor belongs to is also determined. But, from Table 2, we can see that the number of environmental factors in some categories is still big, for example 38 and 34. It is not easy to fully understand a category with so many factors. Therefore, we extracted and named the sub-categories for the eleven categories according to the meanings of factors. The complete listing of the taxonomy is presented in Table 3, which answer the research question RQ3. The numbers in parentheses indicate how many elements are in each category.

<<Table 3>>

In order to provide an easily digestible view of the contents outlined in Table 3, the essential components of the taxonomy of environmental factors influencing individual decision-making are summarized in graphical form in Figure 4. The core of the Figure 3 represents the environmental factors influencing individual decision-making behavior, and eleven categories are represented by rectangles, in which at the top is the name of category, and below are the sub-categories.

Looking at Figure 4, it is easy to understand that these categories will affect individual decision-making behaviors in software projects. Eleven categories are decomposed into 38 sub-categories. Undoubtedly, work itself (the first category), work condition (the third category) and the feedback from the work (the seventh category) influence individual decision-making behaviors. For example, interesting work makes individuals decide to put more time into it. It seems that the fourth and the eighth categories are the same. However, the eighth category addresses organization attributes itself, and the fourth addresses support from the company. Support is very important for individual behaviors, for example, career support from the company allows developers to work more enthusiastically and not easily resign even when facing more pressure from software projects. Goal, the second category in our result, definitely has an impact on individual behaviors, because everyone does something in order to achieve his/her goal. The sixth, ninth, tenth and eleventh categories are closely related to the attributes of software project: teamwork and technology development.

<< Figure 4 >>

**Figure 4** Our taxonomy of environmental factors affecting individual decision-making behaviors in software projects

## 5. Discussions

In this study, we adopted the clustering method to group 237 environmental factors provided in the study[11] into eleven categories, and obtained a relatively objective and comprehensive taxonomy of environmental factors affecting individual decision-making behavior in software projects. In this section, we discuss how the taxonomy can be useful for both the research and practicing communities.

### 5.1 Implications for research

The results of this study have several implications for researchers. First, although there exists much research focusing on how people make decisions, past studies represent a special context which involved limited influential environmental factors. In addition, although a classification of environmental factors affecting individual decision-making behaviors in software projects has been presented in at least one study,[11] we believe that the taxonomy framework presented herein is comprehensive and relatively objective, with appropriate references provided. It is important to profile the important aspects of environmental factors; our taxonomy can help researchers more fully understand the question. Therefore, our categories can provide a base for researchers to develop a better individual decision-making behavior model so as to predict the decision results more accurately.

Second, our dataset (the environmental factors) came from a study[11] that extracted factors by the method of SLR. In general, factors identified by SLR are classified according to the subjective literature analysis, as many studies demonstrate.[6,11,52] In order to overcome the threat of subjectivity of the results, we used clustering technology to classify these factors. Our research extended the application of clustering, and provides a combination of clustering and SLR. It can provide some research insights for SLR.

Third, the measuring similarity of words is based on WordNet. Existing research based on WordNet mostly considers only the synonymy relationship between words, but we also consider the hypernymy between words, and design an algorithm to measure the semantic similarity between words and phrases. The algorithm also provides some insights for researchers who focus on clustering based on WordNet.

Fourth, our categories can support some statements in prior literature. For example, during requirements analysis, support from top management promotes user participation actively.[53] Here, top management support, which belongs to the fourth category, is an external environmental factor that affects the participation behavior of the user. Therefore, our classification provides some insights for researchers to investigate the relationship between external factors and overt behaviors.

### 5.2 Implications for practice

This study looked into the taxonomy of environmental factors affecting individual decision-making behaviors in software projects because of the practical relevance of this issue to industry. Individual decision-making behavior is adaptive, but the environmental factors are complex and diverse. So, the eleven categories in our taxonomy of environmental factors can provide software managers a guideline to recognize which aspects are more important to manage, guide and predict individual decision-making.

Some categories are within the control of managers, for example, the third and the fourth categories (physical conditions and company support). Thus, understanding and managing these environmental factors, which includes creating positive factors or overcoming negative factors, might be significant for managers in order to guide individual decision-making behavior towards the direction of benefiting the project. For example, in order to avoid the core developers' turnover decision, the manager can use the first four categories to keep them, such as providing fair promotion, catering for physical working condition and support, and making the challenging work match their work goal. In addition, Keil et al.[45] have found that developers often make a decision not to report bad news about a project for personal risks, such as losing face that is related to the team, organization and feedback factors. Therefore, in order to learn the real status of a project, manager can advocate and cultivate a positive team and organizational culture: giving positive feedback to those who report bad news. On the other hand, some are beyond the control of managers, for example, personal relationship goal, which account for a considerable proportion of the categories identified. Most of these factors are related to the psychological state of individuals, which in turn can affect individual decisions. For example, different developers have different attitudes toward the factor of balance between personal and professional life. For those who care about this factor, they may be reluctant to and complain of working overtime under time pressure of a project. Therefore, it is important for managers to understand these factors from the viewpoint of psychology, so they can better understand and predict individual behavior; also to further understand the influences of behaviors on a project in order to implement a coping strategy.

In addition, for software development individuals, our categories provide guidance on the environmental factors that may be considered and assessed for a special decision, so individuals can have an environmental awareness and carefully make reasonable decisions. For instance, when estimating effort in agile development, besides the characteristics of the task, developers also may consider the factor of peer commitment for development.

## 6. Limitations and future research

Limitations that pertain to our study need to be acknowledged as results are bounded by threats to validity. The first limitation is related to the values of some parameters in our similarity algorithm. We determined the parameters values based on multiple experiments. However, different parameters may lead to different results, so the parameters values may affect the validity of the results. Therefore, the assessment of the optimal parameters can be carried out in a future study.

Second, our similarity algorithm proposed to group the environmental factors only considers the semantic analysis for short phrases. However, phrases also imply syntax structure, which is ignored in this paper. Thus, there is a possibility that some factors were classified unreasonably. For example, characteristics of the organization is grouped into the category of characteristics of the task, instead of organization. In the future, we should further eliminate this type of limitation using other methods.

Third, although we have grouped 237 environmental factors into 11 categories, we have not associated these categories with different types of decisions and different stages of software projects. After all, for a special decision, its environmental factors can vary; not all categories and environmental factors will have an impact on a special decision. For example, the individual decision-maker about the requirements of software product do not have much direct contact with customers, but the individual decision-maker about the requirements of software applications may communicate with the customers directly. Under these two cases, customer involvement has a different impact on the decision about requirements. Therefore, it is necessary to prioritize the importance of the categories and factors according to the types of individual decision in future research. Additionally, it is worthwhile to identify the relationship between the environmental factors and software development process so as to give the project manager guidance to understand and control the influence of different environmental factors during the different software development stages.

A fourth limitation of the taxonomy we proposed relates to the absence of broad community involvement in the validation of the taxonomy. In order to gain broad consensual agreement for a taxonomy of environmental factors affecting individuals decision-making behaviors in software projects, it require many experts from a broad range of software domains to develop and agree on the constituent factors. In addition, the taxonomy need the practical data support to verify it. So, future detailed studies are necessary of interviewing some professionals in SE domain or observing people in their actual decision scenes to fine tune the taxonomy.

## 7. Conclusions

Individual decision-making behaviors is influenced heavily by external environmental factors. Although environmental factors affecting individual decision-making behaviors in software projects have been identified and classified in prior literature, the method of classifying is subjective, thus, the quality of classification faces the threat of bias. In this study, clustering technique is used to reduce the threat and group these environmental factors. The motivation behind this work is that we believe that clustering analysis can help us more accurately understand which categories of environmental factors exist.

Our aim is to segregate the environmental factors into groups where each group represents some topic that is different than those topics represented by the other groups. Notably, the most challenging concern is how to determine the similarity between two environmental factors to be clustered. Based on synonymy and hypernymy in WordNet, we designed algorithms to measure semantic similarity between words, further phrases (because each environmental factor is a phrase). Then, a k-means analysis was adopted to group these factors. We finally achieved a taxonomy of environmental factors affecting individuals decision-making behaviors in software projects, which include the optimal eleven categories with centroids as the name of category.

Improving decision making in software development hinges on our understanding of decisions. The taxonomy presented herein is quite comprehensive, and provides a first step in this direction by giving an appropriate and relatively objective reference of the environmental factors that affect individual decision-making behaviors in software projects. The optimal number of categories is eleven, which indicates the variety and diversity of environmental factors. Thus, the taxonomy contributes to serving as reminders of the level of complexity of individual decision involved in software projects. Among the eleven categories, apart from team, task, technical and organizational categories, which obviously influence the individual decision-making, we found that goal, support and peer commitment, which are less visible and easily overlooked, also have the influence on individual decision-making behaviors in software projects. The results provide support for practitioners who are challenged with managing a software team. In addition, as mentioned before, the intrinsic personal factors and extrinsic environmental factors are interactive during the decision process. The eleven categories in our taxonomy can abstractly describe and define the external factors, further provide a possibility for researching the interaction between personal and environmental factors in order to better understand individual decision.

**References**
1. Cunha JAOG, Moura HP, Vasconcellos FJS. Decision-making in software project management: a systematic literature review. Procedia Comput Sci 2016;100:947-954.
2. Lenberg P, Feldt R, Wallgren LG. Behavioral software engineering: a definition and systematic literature review. J Syst Softw 2015;107:15-37. doi:10.1016/j.jss.2015.04.084.
3. Alexander PM, Holmner M, Lotriet HH, et al. Factors affecting career choice: comparison between students from computer and other disciplines. J Sci Edu Technol 2011;20:300-315.
4. Lowry PB, Zhang J, Wu T. Nature or nurture? a meta-analysis of the factors that maximize the prediction of digital piracy by using social cognitive theory as a framework. Comput Hum Behav 2017;68:104-120. doi: 10.1016/j.chb.2016.11.015.
5. Ghapanchi AH, Aurum A. Antecedents to IT personnel's intentions to leave: a systematic literature review. J Syst Softw 2011;84:238-249.
6. Clarke P, O'Connor RV. The situational factors that affect the software development process: towards a comprehensive reference framework. Inf Softw Technol 2012;5:433-447.
7. Orvis KA, Leffler GP. Individual and contextual factors: an interactionist approach to understanding employee self-development. Personal Individ Differ 2011;51:172-177.
8. Groher I, Weinreich R. A study on architectural decision-making in context. Proceedings of the 12th Working IEEE/IFIP Conference on Software Architecture, 2015:11-20. doi:10.1109/WICSA.2015.27.
9. Harrison NB, Gubler E, Skinner D. Architectural decision-making in open-source systems -- preliminary observations. Proceedings of the 1st International Workshop on Decision Making in Software ARCHitecture (MARCH), 2016:16-21. doi:10.1109/MARCH.2016.7

10. Cunha JAOGd, Silva FQBd, Moura HPd, Vasconcellos FJS. Towards a substantive theory of decision-making in software project management: preliminary findings from a qualitative study. Proceedings of the 10th ACM/IEEE International Symposium on Empirical Software Engineering and Measurement, Ciudad Real, Spain, 2016:1-10.

11. Jia J, Zhang P, Capretz LF. Environmental factors influencing individual decision-making behavior in software projects: a systematic literature review. Proceedings of the 9th International Workshop on Cooperative and Human Aspects of Software Engineering, Austin, Texas, 2016:86-92.

12. Qimin C, Qiao G, Yongliang W, Xianghua W. Text clustering using VSM with feature clusters. Neural Comput Appl 2015;26:995-1003. doi:10.1007/s00521-014-1792-9.

13. Choi S-J, Kim D-K, Park S. Identifying correlations of findings for building process improvement packages using graph clustering. J Softw: Evol Proc 2015;27:514-527. doi:10.1002/smr.1723.

14. Di Francescomarino C, Marchetto A, Tonella P. Cluster-based modularization of processes recovered from web applications. J Softw: Evol Proc 2013;25:113-138. doi:10.1002/smr.518.

15. LI H. Word clustering and disambiguation based on co-occurrence data. Nat Lang Eng 2002;8:25-42. doi: 10.1017/S1351324902002838.

16. Pérez-Castillo R, Caivano D, Piattini M. Ontology-based similarity applied to business process clustering. J Softw: Evol Proc 2014;26:1128-1149. doi:10.1002/smr.1652.

17. Hammouda KM, Kamel MS. Efficient phrase-based document indexing for Web document clustering. IEEE T Knowl Data En 2004;16:1279-1296. doi:10.1109/TKDE.2004.58.

18. Bekkerman R, El-Yaniv R, Tishby N, Winter Y. Distributional word clusters vs. words for text categorization. J Mach Learn Res 2003;3:1183-1208.

19. Chen C-L, Tseng FSC, Liang T. An integration of WordNet and fuzzy association rule mining for multi-label document clustering. Data Knowl Eng 2010;69:1208-1226.

20. Bassiou N, Kotropoulos C. Long distance bigram models applied to word clustering. Pattern Recogn. 2011;44:145-158. doi: 10.1016/j.patcog.2010.07.006.

21. Li Y, Chung SM, Holt JD. Text document clustering based on frequent word meaning sequences. Data Knowl Eng 2008;64:381-404. doi: 10.1016/j.datak.2007.08.001.

22. Zheng H-T, Kang B-Y, Kim H-G. Exploiting noun phrases and semantic relationships for text document clustering. Inform Sciences 2009;179:2249-2262. doi: 10.1016/j.ins.2009.02.019.

23. Wang P, Xu B, Xu J, et al. Semantic expansion using word embedding clustering and convolutional neural network for improving short text classification. Neurocomputing 2016;174, Part B: 806-814.

24. Pera MS, Yiu-Kai N. Utilizing phrase-similarity measures for detecting and clustering informative RSS news articles. Integr Comput-Aid Eng 2008;15:331-350.

25. Momtazi S, Klakow D. A word clustering approach for language model-based sentence retrieval in question answering systems. In Proceedings of the 18th ACM conference on Information and knowledge management, Hong Kong, China, 2009:1911-1914. doi:10.1145/1645953.1646263.

26. Hackenberg M, Rueda A, Carpena P, Bernaola-Galván P, Barturen G, Oliver JL. Clustering of DNA words and biological function: a proof of principle. J Theor Biol 2012;297:127-136.

27. Guo X, Yu Q, Alm CO, Calvelli C, Pelz JB, Shi P, Haake AR. From spoken narratives to domain knowledge: mining linguistic data for medical image understanding. Artif Intell Med 2014;62:79-90.

28. Wu Y-C. A top-down information theoretic word clustering algorithm for phrase recognition. Inform Sciences 2014;275:213-225. doi: 10.1016/j.ins.2014.02.033.

29. Chen Q, Chen Y, Jiang M. Cluster analysis based on contextual features extraction for conversational corpus. J Comput Commun 2015;3:33-37.
30. Wei T, Lu Y, Chang H, Zhou Q, Bao X. A semantic approach for text clustering using WordNet and lexical chains. Expert Syst Appl 2015;42(4):2264-2275.
31. Arora P, Deepali, Varshney S. Analysis of K-Means and K-Medoids algorithm for Big Data. Procedia Comput Sci 2016;78:507-512. doi: 10.1016/j.procs.2016.02.095.
32. Wang X, Bai Y. The global minmax k-means algorithm. SpringerPlus 2016;5(1):1665.
33. Bouras C, Tsogkas V. A clustering technique for news articles using WordNet. Knowl-Based Syst 2012;36:115-128. doi: 10.1016/j.knosys.2012.06.015.
34. Lee S, Huh S-Y, McNiel RD. Automatic generation of concept hierarchies using WordNet. Expert Syst Appl 2008;35(3):1132-1144. doi: 10.1016/j.eswa.2007.08.042.
35. Geum Y, Park Y. How to generate creative ideas for innovation: a hybrid approach of WordNet and morphological analysis. Technol Forecast Soc Chang 2016;111:176-187.
36. Zhu G, Iglesias CA. Computing semantic similarity of concepts in knowledge graphs. IEEE T Knowl Data Eng 2017;29:72-85.
37. Mukherjee N, Dicks LV, Shackelford GE, Vira B, Sutherland WJ (2016) Comparing groups versus individuals in decision making: a systematic review protocol. Environ Evid 5 (1):19. doi:10.1186/s13750-016-0066-7
38. Tessem B. Individual empowerment of agile and non-agile software developers in small teams. Inf Softw Technol 2014;56:873-889. doi: 10.1016/j.infsof.2014.02.005.
39. Alenljung B, Persson A. Portraying the practice of decision-making in requirements engineering: a case of large scale bespoke development. Requirements Eng 2008;13:257-279.
40. Aurum A, Wohlin C. The fundamental nature of requirements engineering activities as a decision-making process. Inf Softw Technol 2003;45:945-954. doi: 10.1016/S0950-5849(03)00096-X.
41. Zannier C, Chiasson M, Maurer F. A model of design decision making based on empirical results of interviews with software designers. Inf Softw Technol 2007; 49:637-653.
42. Zannier C, Maurer F. Comparing decision making in agile and non-agile software organizations. Proceedings of the 8th international conference on Agile processes in software engineering and extreme programming, Como, Italy2007;1-8.
43. Drury M, Conboy K, Power K. Obstacles to decision making in agile software development teams. J Syst Softw 2012;85:1239-1254. doi: 10.1016/j.jss.2012.01.058.
44. Børte K, Ludvigsen SR, Mørch AI. The role of social interaction in software effort estimation: unpacking the "magic step" between reasoning and decision-making. Inf Softw Technol 2012; 54:985-996. doi: 10.1016/j.infsof.2012.03.002
45. Keil M, Im GP, Mähring M. Reporting bad news on software projects: the effects of culturally constituted views of face-saving. Inform Syst J 2007;17:59-87.
46. Hahn J, Moon JY, Zhang C. Emergence of new project teams from open source software developer networks: impact of prior collaboration ties. Inform Syst Res 2008;19:369-391.
47. Sharma P, Savarimuthu BTR, Stanger N, et al. Investigating developers' email discussions during decision-making in Python language evolution. Proceedings of the 21st International Conference on Evaluation and Assessment in Software Engineering, Karlskrona, Sweden, 2017:286-291.

48. Hirao T, Ihara A, Matsumoto K-i. Pilot study of collective decision-making in the code review process. Proceedings of the 25th Annual International Conference on Computer Science and Software Engineering, Markham, Canada, 2015:248-251.

49. Mesh ES, Tolar DM, Hawker JS. Exploring process improvement decisions to support a rapidly evolving developer base. Proceedings of the IEEE/ACM 38th International Conference on Software Engineering Companion (ICSE-C), Austin, TX, USA, 2016:777-780. doi: 10.1145/2889160.2889209.

50. Nakatani T, Koiso Y. A method for analyzing the context of stakeholders and their requirements. Proceedings of the 9th International Conference on Software Engineering and Applications (ICSOFT-EA), Vienna, Austria, 2014:357-362.

51. Capó M, Pérez A, Lozano JA. An efficient approximation to the K-means clustering for massive data. Knowl-Based Syst 2017;117:56-69. doi:10.1016/j.knosys.2016.06.031.

52. Thuan NH, Antunes P, Johnstone D. Factors influencing the decision to crowdsource: a systematic literature review. Inform Syst Front 2016;18(1):47-68. doi:10.1007/s10796-015-9578-x.

53. Jia J, Capretz LF. Direct and mediating influences of user-developer perception gaps in requirements understanding on user participation. Requirements Eng 2017; available online. doi:10.1007/s00766-017-0266-x.

Table 1 The values of three kinds of semantic similarities under different $k$

| The number of categories ($k$) | the maximum semantic similarity ($S_{kmax}$) | the minimum semantic similarity ($S_{kmin}$) | the average semantic similarity ($S_{kavg}$) |
|---|---|---|---|
| 5 | 0.3437 | 0.1990 | 0.2778 |
| 6 | 0.3408 | 0.2469 | 0.2873 |
| 7 | 0.4509 | 0.2110 | 0.3057 |
| 8 | 0.4592 | 0.2684 | 0.3577 |
| 9 | 0.4931 | 0.2710 | 0.3723 |
| 10 | 0.8 | 0.2557 | 0.4344 |
| 11 | 0.79 | 0.2896 | 0.4404 |
| 12 | 0.7409 | 0.2483 | 0.4457 |

Table 2 the categories of environmental factors

| No. | Category name | The number of factors |
|---|---|---|
| 1 | Challenging work | 38 |
| 2 | Goal | 34 |
| 3 | Appropriate physical conditions | 24 |
| 4 | Company support | 23 |
| 5 | *Characteristics of the task* | 20 |
| 6 | *Distributed team* | 20 |
| 7 | Feedback from the job | 20 |
| 8 | *Organization* | 20 |
| 9 | *Technical competence* | 18 |
| 10 | Development | 11 |
| 11 | Peer commitment | 9 |

Table 3 The list of the classification

| Sub-categories | factors |
|---|---|
| *Challenging work (38)* | |
| Work characteristic (14) | Challenging work/ collaborative work/ interesting work/ technically challenging work/ the work/ variety of work / work and life balance/ work conditions/ work environment/ work environment flexibility/ work lifestyle/ work-personal life balance/ quality of work/ quantity of work |
| Participation (2) | Customer involvement/ employee participation |
| Benefit (8) | Benefits/ competitive salary/ justifiable benefits/ non-financial benefits/ pay/ payment/ salary/ uncompetitive salary |
| Promotion (2) | Lack of promotion/ promotion opportunities |
| Work practice (12) | Compulsory/ communication effectiveness/ experimentation/ reduced admin/ shared best practices/ strong work ethic/ technology to help work/ quality of work performed/ quantity of work performed/ maintainable/ work accomplishment/ work management |
| *Goal (34)* | |
| Goal attribute (11) | Goal/ goal acceptance/ goal achievement/ goal clarity/ goal difficulty/ goal setting participation/ goals and priorities based on non-technical/ political goal/ politically driven goals/ unrealistic goals/ type |
| Soft goal (9) | Broad personal skill/ cohesion/ cohesion and synergy aspects/ collaboration/ creativity/ opportunities/ quality of management/ recognition/ meeting targets/ |
| Goal condition (6) | Infrastructure/ working infrastructure/ good management/ skill variety/ poor culture fit /resources management |
| Personal relation goal (4) | Interpersonal relations with subordinate/ interpersonal relations with superior/ recognitions from others/ lack of influence |
| Product (4) | Customer expectations/ meaningful products/ nature of products/ perceived value/ |
| *Peer commitment (9)* | |
| Peer's support (6) | Commitment/ interpersonal relation with peers/ peer commitment/ peer motivation/ peer proactivity/ peer trustworthiness |
| Other's commitment (3) | Customer confidence/ organizational commitment/ top-down commitment |
| *Appropriate physical conditions (24)* | |
| Physical condition (5) | Appropriate physical conditions/ appropriate technological conditions/ appropriate working conditions/ workload/ positive effect on team-work conditions |
| Soft condition (13) | Autonomy/ balance between personal and professional life/ career progression opportunities/ personal life/ professionalism/ self-efficacy/ standardization/ status/ threats of punishment/ customer satisfaction/ inequity/ trust and worthiness/ trust and respect |
| Challenge (3) | Challenge/ intellectual challenge/ technological challenge |
| Risk (3) | Degree of risk/ risk/ risks reassessed and controlled |
| *Development (11)* | |
| Technology development (6) | Development/ development needs addressed/ development practices/ phrased development/ software development factors/ technical development |
| Soft development (5) | Career development/ people development/ performance improvements/skill development/ sales and marketing |
| *Company support (23)* | |
| Career support (3) | Career path/ career planning support/ career progression support |
| Company success (3) | Successful company/ visible success/ working in successful companies |
| Implicit support factor (10) | Bureaucracy/ change/ company policy and administration/ cost beneficial/ culture/ fun/ integration/ eliminates bureaucracy/ national culture/ staff turnover |
| Company support (6) | Collaborative support/ customer support/ management support/ technical support/ support/ company support |
| Customer support (1) | Customer support |
| *Distributed teams (20)* | |
| Team staff (4) | Adequate staff/ characteristics of team members/ knowledgeable team leaders/ number of employees |
| Team spirit (6) | Bottom-up initiatives/ sense of belonging/ staff appreciation/ team cohesion/ team motivation/ learning |
| Team resources (7) | Budgets/ distributed team/ process ownership/ task significance/ team quality/ team size/ team working |
| Team ability (3) | Intellectual problem solving/ managerial practice/ problems resolution |
| *Feedback from the job (20)* | |
| Direct feedback (6) | Customer feedback/ feedback/ feedback from supervisors/ feedback from the job/ feedback on goal accomplishment/ managerial feedback |
| Feedback factor (9) | Business/ empowerment/ learning opportunities/ nature of business/ participation in the entire life cycle of a project/ planning effectiveness/ project management factors/ project delivery time/ project success |
| Job feedback (5) | Job outcomes/ job performance/ job security/ job stability/ job satisfaction |
| *Organization (20)* | |
| Organization attribute (6) | Organization/ organization size/ organizational contexts/ organizational factors/ organizational practices/ years of organizational existence |
| Reward (4) | Reward schemes/ reward system/ rewards and financial incentives/ staff rewards |
| Time and stress (6) | Managing time and stress/ stress/ time/ time-pressure/ well-defined work-time/ work-time flexibility |
| Organizational practices (4) | Participation in decision making/ recruiting and selection/ training/ obstacles |
| *Technical competence (18)* | |
| Competence (4) | Competence/ peer technical competence/ technical competence/ technical supervision |
| Factors influencing technical competence (8) | Decision making/ equity/ external audits/ motivating and influencing/ motivation/ resources/ sufficient resources/ leadership influencing |
| Software quality (3) | Producing poor quality software/ quality of software/ quality software |
| User relationship (3) | Relationship opportunities/ relationships with users/ user relationship |
| *Characteristics of the task (20)* | |
| Task characteristics (9) | Task characteristics/ task complexity/ task forces/ task identity/ task variety/ characteristics of task/ refactoring as assigned tasks/ flexibility/ production |
| Project manager (3) | Project manager authority/ project manager communication/ project manager vision |
| Project (3) | Project outcome/ project size/ external project factors |
| Soft factors related to tasks (5) | Communication/ characteristics of the organization/ external opportunities/ teammates characteristics/ time and task management |